\input epsf 
\documentclass[preprint,eqsecnum,aps]{revtex4}
\usepackage{graphicx}
\def\beq{\begin{eqnarray}}
\def\eeq{\end{eqnarray}}

\begin{document}
\title{Presenting a new method for the solution of nonlinear problems}
\author{Paolo Amore~\footnote{paolo@ucol.mx} and Alfredo Aranda~\footnote{fefo@cgic.ucol.mx}}
\affiliation{Facultad de Ciencias, Universidad de Colima, Colima, M\'exico}
\date{March 17, 2003}
\vskip .1in
\thispagestyle{empty}

\begin{abstract}    

We present a method for the resolution of (oscillatory) nonlinear problems. It is based on the
application of the Linear Delta Expansion to the Lindstedt-Poincar\'e method. By applying it to the
Duffing equation, we show that our method substantially improves the approximation given by the simple 
Lindstedt-Poincar\'e method.

\end{abstract}                                                

\maketitle

\section{Introduction}
\label{sec:intro}
In this letter we present a method for the resolution of oscillatory nonlinear problems. 
This new method is obtained by applying the Linear Delta Expansion (LDE)~\cite{lde} to the well-known 
Lindstedt-Poincar\'e (LP) method~\cite{Lin83}. We explicitly solve the Duffing Equation (Anharmonic Oscillator
Equation) and find that our approximation works much better over a wider range of parameters than does
the simple LP method. We also compare our results to an approximation performed using the
method of the perturbative $\delta$ expansion~\cite{BMPS89}, and find again that our approximation 
works with better accuracy and convergence for a wider range of parameters. 

In Section~\ref{sec:LPLDE} we briefly review the LP and LDE approximation methods, then in Section~\ref{sec:AO} we show
how by applying the LDE to the LP method, one can solve the Duffing Equation. Our results are presented
in Section~\ref{sec:results} and we finally present our conclusions and directions for future work in 
Section~\ref{sec:conclusions}.

\section{Approximation Schemes}
\label{sec:LPLDE}
\subsection{The Lindstedt-Poincar\'e method}
In this section we introduce the  Lindstedt-Poincar\'e distorted time (LP) method \cite{Lin83}. 
We consider a nonlinear  ODE of the form
\beq
\ddot{x}(t) + \omega^2 \ x(t) = \varepsilon \ f(x(t)) \ ,
\label{LP1}
\eeq
which describes a conservative system, oscillating with an unknown period $T$. The nonlinear term
$\varepsilon \ f(x(t))$ is treated as a perturbation. Unfortunately, when the ordinary perturbation is applied 
to eq. (\ref{LP1}), by writing the solution as a series in $\varepsilon$, the appearance of secular terms spoils 
the expansion and any predictive power is lost for sufficiently large time scales.

In order to avoid the appearance of secular terms, we switch to a scaled time $\tau = 2 \pi t/T \equiv \Omega \ t$, 
where $T$ is the (unknown) period of the oscillations. The ODE now reads:
\beq
\Omega^2 \frac{d^2x}{d\tau^2}(\tau) + \omega^2 \ x(\tau) = \varepsilon \ f(x(\tau)) \ .
\label{LP2}
\eeq

We notice that the dependence upon $\varepsilon$ in this equation enters both in the solution $x(\tau)$ and in the 
frequency $\Omega$. By assuming $\varepsilon$ to be a small parameter we write 
\beq
\Omega^2 &=& \sum_{n=0}^{\infty} \ \varepsilon^n \ \alpha_n \ \ \ ; \ \ \
x(\tau) = \sum_{n=0}^{\infty} \ \varepsilon^n \ x_n(\tau) \nonumber 
\eeq
and expand the r.h.s of eq. (\ref{LP2}) as
\beq
f(x) &=& f\left( \sum_{n=0}^{\infty} \ \varepsilon^n \ x_n(\tau) \right) \approx 
f(x_0) + \varepsilon \ x_1 \ f'(x_0) + \varepsilon^2 \ \left[ x_2 \ f'(x_0) + \frac{x_1^2}{2} \ f''(x_0) \right] \nonumber \\
&+& \varepsilon^3 \ \left[ x_3 \ f'(x_0) +  x_2 \ x_1 \  f''(x_0) + \frac{x_1^3}{6} \ f'''(x_0)
 \right] + O\left[\varepsilon^4\right] \nonumber \ .
\eeq

By using these expansions inside  eq.~(\ref{LP2}) we obtain a system of linear inhomogeneous differential equations, 
each corresponding to a different order in $\varepsilon$. Let us consider the first few terms. 
To order $\varepsilon^0$ we obtain the equation
\beq
\alpha_0 \ \frac{d^2x_0}{d\tau^2} + \omega^2 \ x_0(\tau)  &=& 0 \ ,
\label{LP3}
\eeq
describing a harmonic oscillator of frequency $\Omega = \sqrt{\alpha_0} =\omega$.  
To order $\varepsilon$ we obtain the equation
\beq
\alpha_0 \ \frac{d^2x_1}{d\tau^2} + \omega^2 \ x_1(\tau)  = s_1(\tau) \ ,
\label{LP4}
\eeq
where the r.h.s. is given by
\beq
s_1(\tau) &\equiv& - \alpha_1 \  \frac{d^2x_0}{d\tau^2} + f(x_0) \ .
\eeq

We remark the oscillatory behavior of the driving term $s_1(\tau)$, because of its dependence upon the 
order-0 solution, $x_0(\tau)$. 
As a result $s_1(\tau)$ will contain the fundamental frequency, corresponding to a period of $2 \pi$ in the scaled time, and 
multiples of this frequency, appearing through the term $f(x_0(\tau))$. The presence of a driving term with the 
fundamental frequency leads to a resonant behavior of $x_1(\tau)$ and to the unfortunate occurrence of secular terms, which
spoils our expansion. However, we can deal with this problem by fixing the coefficient $\alpha_1$ to cancel the resonant 
term in the r.h.s. of eq. (\ref{LP4}). The iteration of this procedure to a given order $n$ allows to determine
the coefficients
$\alpha_0, \dots, \alpha_n$ and therefore the frequency $\Omega = \sqrt{\alpha_0 + \alpha_1 + \dots + \alpha_n}$.

\subsection{Linear delta expansion}

The linear delta expansion (LDE) is a powerful technique which has been originally introduced 
to deal with problems of strong coupling Quantum Field Theory, for which the naive perturbative approach is not 
useful. Since then this method has been applied to a wide class of 
problems~\cite{Jones:1991vu,blencowe,Kneur:2002dn,Kneur:2002kq,Krein:1995rp,Pinto:1999py}.
In its original formulation a lagrangian density ${\cal L}$, which is not exactly solvable, is interpolated with
a solvable lagrangian ${\cal L}_0(\mu)$, depending upon one (or more) parameters $\mu$:
\beq
{\cal L}_\delta = {\cal L}_0(\mu) + \delta \ \left( {\cal L} - {\cal L}_0(\mu) \right) \ .
\eeq

For $\delta=0$ one obtains ${\cal L}_0(\mu)$, whereas for  $\delta=1$ one recovers the full lagrangian ${\cal L}_\delta$. 
The term $\delta \ \left( {\cal L} - {\cal L}_0 \right)$ is treated as a perturbation and $\delta$
is used to keep track of the perturbative order. Eventually $\delta$ is set to be $1$. 

We notice that the interpolation of the full lagrangian with the solvable one, ${\cal L}_0(\mu)$, brings an artificial 
dependence upon the arbitrary parameter $\mu$. Such dependence, which would vanish if all perturbative orders 
were calculated,
can be milden to a finite perturbative order, by requiring some physical observable $\cal O$ to be locally insensitive to 
$\mu$, i.e:
\beq
\frac{\partial{\cal O}(\mu)}{\partial \mu} = 0 \nonumber .
\eeq
This condition is known as Principle of Minimal Sensitivity (PMS) and is normally 
seen to improve the convergence to the exact 
solution.

\section{Anharmonic oscillator}
\label{sec:AO}

Following the techniques described in the previous Section, we can now apply the LDE to the LP for
the solution of the Duffing Equation (the Anharmonic oscillator equation):
\beq
\frac{d^2x}{dt^2}(t) + \omega^2 \ \ x(t) = - \mu \ x^3(t) \ .
\label{duf1}
\eeq
This equation describes a conservative system, where the total energy is
given by 
\beq
E = \frac{\dot{x}^2}{2} + \left[ \frac{\omega^2 \ x^2}{2} + \mu \ \frac{x^4}{4} \right] \ .
\eeq
The period of the oscillation can be calculated in terms of an elliptic integral
\beq
T_{exact} &=& 2 \ \int_{-A}^{A} dx \ \frac{1}{\sqrt{{2 (E-V(x))}}} \ ,
\label{periodex}
\eeq
where $A$ is the amplitude of the oscillations.

Following the procedure explained in the previous Section, we write
Eq.~(\ref{duf1}) as
\beq
\Omega^2 \frac{d^2x}{d\tau^2}(\tau) + \left( \omega^2 + \lambda^2\right) \ x(\tau) = \delta \left[
- \mu \ x^3(\tau) + \lambda^2  \ x(\tau)\right]  \ ,
\label{duf2}
\eeq
where an arbitrary parameter $\lambda$ with dimension of frequency has been introduced. 
Clearly for $\delta = 1$, Eq.~(\ref{duf2}) reduces to Eq.~(\ref{duf1}). We repeat the procedures previously 
explained and find a hierarchy of linear inhomogeneous differential equations 
to be solved sequentially. 

To zeroth order we obtain the equation
\beq
\alpha_0 \ \frac{d^2x_0}{d\tau^2} + (\omega^2+\lambda^2) \ x_0(\tau)  &=& 0 \ ,
\eeq
with solution
\beq
x_0(\tau) &=& A  \ \cos\tau \ .
\eeq

The zeroth order frequency is then given by
\beq
\alpha_0 &=& \omega^2 +  \lambda^2  \ .
\eeq

We proceed to compute the first order and find that 
\beq
\alpha_0 \ \frac{d^2x_1}{d\tau^2} + (\omega^2+\lambda^2) \ x_1(\tau)  
          &=& S_1(\tau) \, ,
\eeq
where
\beq
S_1(\tau) &=& A \ \cos\tau 
\left[ \alpha_1 + \lambda^2 - \frac{3 A^2 \mu}{4} \right] - \frac{A^3 \mu}{4} \ \cos 3\tau \ .
\eeq
Now $\alpha_1$ is fixed by eliminating the term proportional to $\cos\tau$:
\beq
\alpha_1 &=& \frac{3 A^2 \mu}{4} - \lambda^2 \ .
\eeq

We obtain the solution 
\beq
x_1(\tau) &=& - \frac{A^3 \ \mu}{ 32 (\omega^2+\lambda^2)} \ \cos\tau+ 
 \frac{A^3 \ \mu}{ 32 (\omega^2+\lambda^2)} \ \cos 3\tau  \nonumber \ ,
\eeq
and the frequency
\beq
\Omega^2 = \alpha_0 +\alpha_1 = \omega^2 + \frac{3 A^2 \mu}{4} \, ,
\eeq
which is observed to be  independent of $\lambda$.

The next order gives: 
\beq
\alpha_0 \ \frac{d^2x_2}{d\tau^2} + (\omega^2+\lambda^2) \ x_2(\tau) 
&=& S_2(\tau) \, ,
\eeq
where now
\beq
S_2(\tau)
&=&
\frac{A \left( 3 \ A^4 \ \mu^2+128 \ \alpha_2 \ (\omega^2+\lambda^2) \right)}{128
 \ (\omega^2+\lambda^2)} \ \cos\tau  \nonumber \\
&+& \frac{A^3 \ \mu \ (3 \ A^2 \ \mu - 4 \ \lambda^2)}{16 \  (\omega^2+\lambda^2) } 
\ \cos 3\tau \\ \nonumber
&-& \frac{3 \ A^5 \ \mu^2}{128 \  (\omega^2+\lambda^2)} \ \cos 5\tau \, .
\eeq
As before $\alpha_2$ is fixed by eliminating the term proportional to $\cos\tau$:
\beq
\alpha_2 &=& - \frac{3 \ A^4 \ \mu^2}{128 \  (\omega^2+\lambda^2)} \, .
\eeq

We obtain the solution
\beq
x_2(\tau) &=& \frac{A^3 \mu \ (23 A^2 \mu-32 \lambda^2)}{1024   (\omega^2+\lambda^2)^2} \ \cos\tau + 
\frac{A^3 \mu (- 3 A^2 \mu+ 4 \lambda^2)}{128  (\omega^2+\lambda^2)^2}  \ \cos 3\tau \nonumber \\
&+& \frac{A^5 \mu^2}{1024  (\omega^2+\lambda^2)^2}  \ \cos 5\tau
\eeq
and the frequency
\beq
\Omega^2 = \alpha_0 +\alpha_1 + \alpha_2 = \omega^2 + \frac{3 A^2 \mu}{4} -
\frac{3 A^4 \mu^2}{128 \  (\omega^2+\lambda^2)}  \ .
\eeq
Note that at this order the frequency now depends on the arbitrary parameter $\lambda$. However, due to the
explicit dependence, by applying the PMS, we would obtain the same solution as in the simple LP method. 
In order to get a different solution, we must go to the next order in the expansion.

Finally, following the same procedure, we obtain the expression for the third order:
\beq
\alpha_0 \ \frac{d^2x_3}{d\tau^2} &+& (\omega^2+\lambda^2) \ x_3(\tau) 
=  S_3(\tau) \, ,
\eeq
where
\beq
s_3(\tau) 
&=& \left[ A \ \alpha_3 - 
\frac{3 \ A^5 \ \mu^2 \ (3 \ A^2 \ \mu - 4 \ \lambda^2)}{512 \ (\omega^2+\lambda^2)^2} \right] \ 
\cos\tau \nonumber \\
&-& \frac{(A^3 \ \mu \ (297 \ A^4 \ \mu^2 - 768 \ A^2 \ \mu \ \lambda^2 + 512 \ 
\lambda^4))}{2048 \ (\omega^2+\lambda^2)^2}  \ 
\cos 3\tau  \nonumber \\ 
&+& \frac{3 \ A^5 \ \mu^2 \ (3 \  A^2 \ \mu-4 \ \lambda^2)}{256 \ (\lambda^2 + \omega^2)^2}
 \ \cos 5\tau - \frac{3 \ A^7 \ \mu^3}{2048 \ (\lambda^2+\omega^2)^2} \ \cos 7\tau \, .
\eeq

By eliminating the term proportional to $\cos\tau$ we determine $\alpha_3$ to be 
\beq
\alpha_3 &=& \frac{3 \ A^4 \ \mu^2 (3 \ A^2 \ \mu - 4 \ \lambda^2)}{512 \ (\lambda^2+\omega^2)^2} \, ,
\eeq
and the solution 
\beq
x_3(\tau) &=& - \frac{A^3 \ \mu}{32768} \ \frac{547 \ A^4 \ \mu^2 - 1472 \ A^2 \ \mu \ \lambda^2 + 1024 \ 
\lambda^4}{(\lambda^2+\omega^2)^3} \ \cos\tau \nonumber \\
&+& \frac{A^3 \ \mu}{16384} \ \frac{297 \ A^4 \ \mu^2 - 768 \ A^2 \ \mu \ \lambda^2 + 512 \ 
\lambda^4}{(\lambda^2+\omega^2)^3} \ \cos 3\tau \nonumber \\
&+& \frac{A^5 \ \mu^2}{2048} \ \frac{(-3 \ A^2 \ \mu + 4 \ \lambda^2)}{(\lambda^2+\omega^2)^3} \ \cos 5\tau 
+ \frac{A^7 \ \mu^3}{32768} \ \frac{1}{(\lambda^2+\omega^2)^3} \ \cos 7\tau \, .
\nonumber 
\eeq

The frequency to order $\delta^3$ is now obtained to be
\beq
\Omega^2 = \alpha_0 +\alpha_1 + \alpha_2 + \alpha_3 = 
\omega^2 + \frac{3 A^2 \mu}{4}  -
\frac{3 \ A^4 \ \mu^2}{128 \  (\omega^2+\lambda^2)} +
\frac{3 \ A^4 \ \mu^2 (3 \ A^2 \ \mu - 4 \lambda^2)}{512 \ (\lambda^2+\omega^2)^2} \, .
\label{omega3duff0}
\eeq
This time, the frequency depends upon the arbitrary parameter $\lambda$ in a nontrivial way and
we can apply the PMS in order to fix the value of $\lambda$. We do this by 
imposing that $\frac{d\Omega^2}{d\lambda} = 0$, which leads to the following result:
\beq
\lambda &=& A \ \frac{\sqrt{3 \ \mu}}{2} \, .
\eeq
Notice that since $\lambda$ depends linearly upon $A$ the formula for $\Omega^2$ obtained in this case 
{\sl does not simply correspond to an expansion in $A$}.
As a matter of fact we find that the frequency corresponding to this value of $\lambda$ is
\beq
\Omega^2 = \frac{64 \ A^4 \ \mu^2+192 \ A^2 \ \mu \ \omega^2 + 128 \ \omega^4}{96 \ A^2 \ \mu + 128 \ \omega^2} \, .
\label{omega3duff}
\eeq

Notice that the Duffing equation (\ref{duf1}) is left invariant under the simultaneous rescaling of the anharmonic 
coupling  $\mu$ and of the amplitude, i.e. $\mu \rightarrow \mu'$ and $A \rightarrow A' = A \ \sqrt{\mu/\mu'}$. This
invariance is manifest in the equation (\ref{omega3duff}), which is function of $A^2 \ \mu$, which is invariant under
this rescaling.

\section{Results}
\label{sec:results}
We now present the results following from this analysis. In Fig.~1 we compare 
the exact frequency, calculated with Eq.~(\ref{periodex}) 
with the frequency obtained with our method (LPLDE),
equation (\ref{omega3duff}), and with the LP method, equation (\ref{omega3duff0}) taking $\lambda=0$,
both  to third order in perturbation theory. We take $\omega = \mu = 1$ and vary the amplitude of the oscillations. 
We observe that our method yields an excellent approximation to the exact result even for large amplitudes, 
where the simple LP approximation fails. 

In Fig.~2 we compare the period obtained with our method to the exact period of Eq.~(\ref{periodex}) and
to the one obtained with the formulae of \cite{BMPS89}, which are obtained by applying the nonlinear delta expansion.
Our method provides an excellent approximation to the exact period over a wide range of 
the parameter $\mu$, which controls the 
nonlinearity. The plots are obtained assuming $\omega = 1$ and the boundary conditions $x(0) = 1$ and $\dot{x}(0) = 0$.
The formulae of \cite{BMPS89}  behave badly in the region $\mu < 0$, which corresponds to a potential well of
finite depth centered around $x = 0$, and yield a precision comparable to the one achieved with our method for $\mu >0$.  
Corresponding to the value $\mu = 0$ the oscillator is in a position of (unstable) equilibrium 
and the exact period diverges. Notice that for large values of $\mu$ all the methods seem to give a good approximation
to the exact solution, including the LP method (to first order), which (to third order) was behaving poorly in the case 
previously studied. 
Unfortunately the equations of \cite{BMPS89} are not suitable to be analyzed as in 
Fig.~1, and thus we cannot fully test
the efficiency of this method.

In Fig.~3 we plot the relative error corresponding to the different approximations for $\mu >0$. Our method
to third order in perturbation theory yields an error typically smaller than the errors of the other methods and with a 
magnitude of about $0.1 \ \%$. 

In Fig.~3 we plot the relative error corresponding to the different approximations for $\mu >0$. Our method
to third order in perturbation theory yields an error typically smaller than the errors of the other methods and with a 
magnitude of about $0.1 \ \%$. 

\section{Conclusions}
\label{sec:conclusions}
We have presented a method for the solution of (oscillatory) nonlinear problems.
It is based on the application of the Linear Delta Expansion to the Lindstedt-Poincar\'e method. 
We applied it to the Duffing Equation and find that the 
new method converges faster and with greater accuracy than the simple LP method. Also, by comparing it
with methods based on the perturbative $\delta$ expansion, we show that our solution not only converges faster and
more accurately, but it also works for a much wider range of parameters. We are currently applying
our method to a wider class of nonlinear problems~\cite{WP}, and we are also interested in considering its possible
extension to quantum systems.

\section{Acknowledgments}

The Authors acknowledge the support of the ``Fondo Alvarez-Buylla'' of the University of Colima
and of Conacyt in the completion of this work.


\newpage


\newpage

\begin{figure}
\includegraphics[width=12cm]{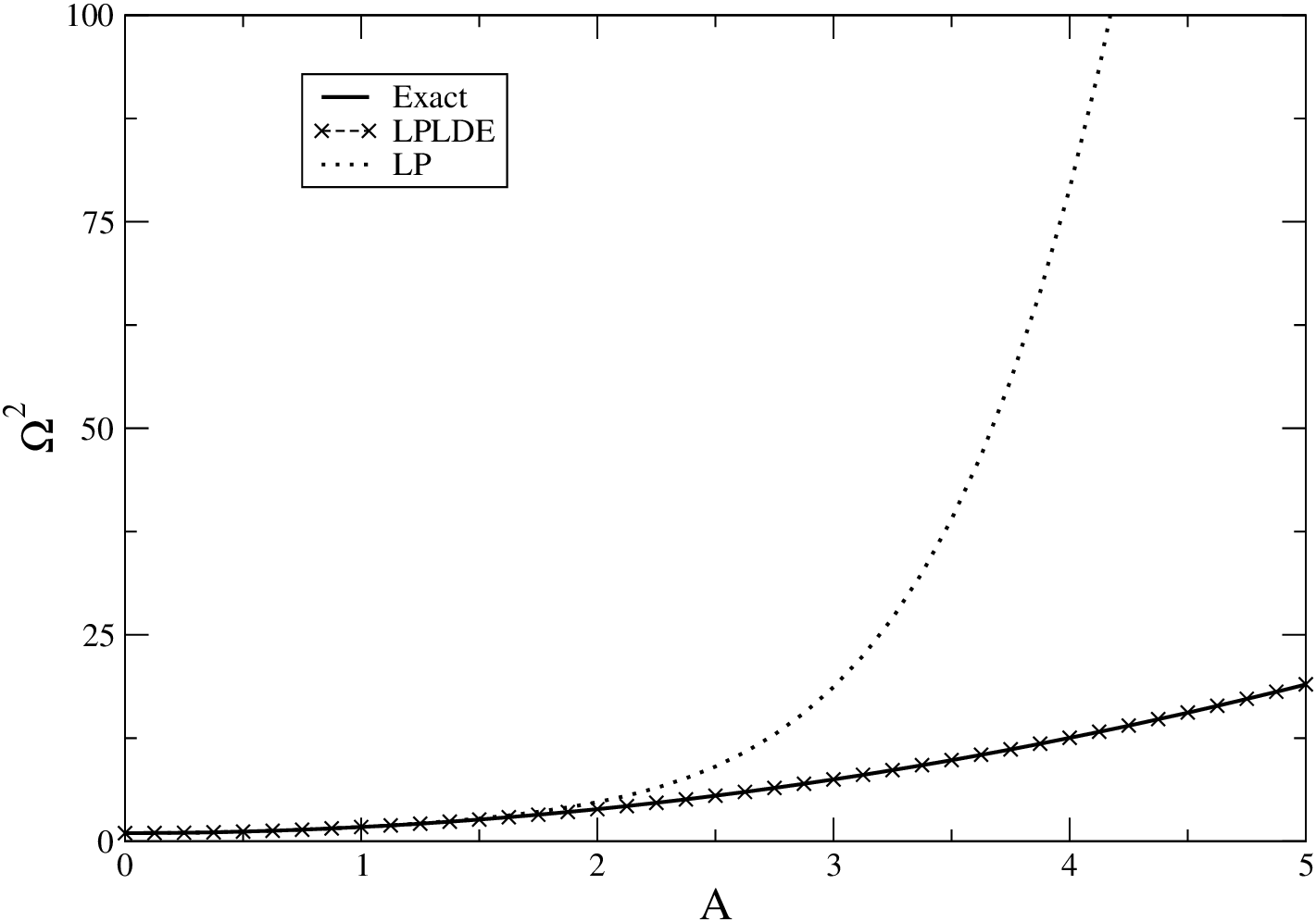}
\caption{Squared frequency of the anharmonic oscillator as a function of the amplitude 
(arbitrary units). $\omega = \mu =1$.}
\end{figure}

\begin{figure}
\includegraphics[width=12cm]{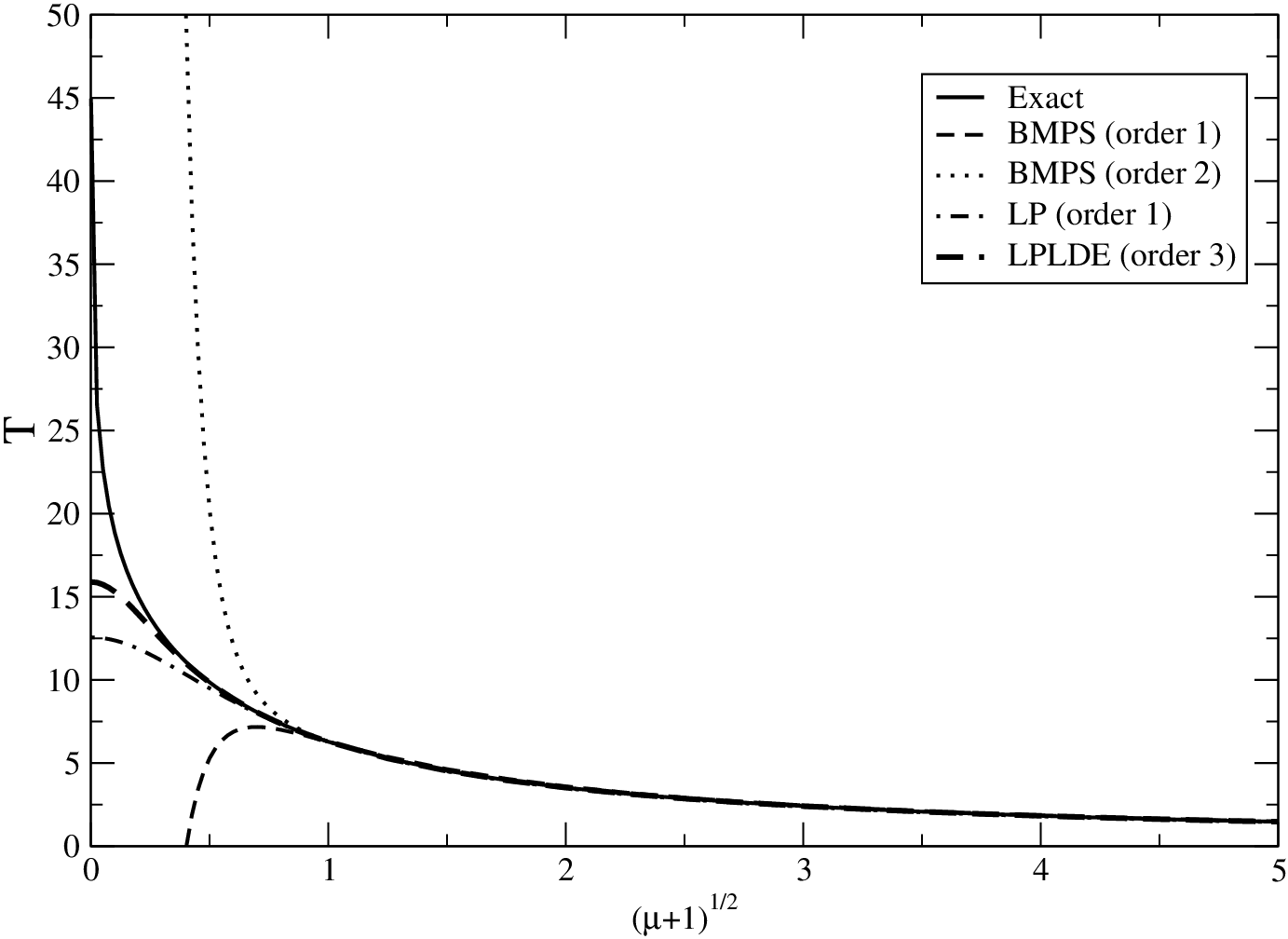}
\caption{Period of the anharmonic oscillator.}
\end{figure}

\begin{figure}
\includegraphics[width=12cm]{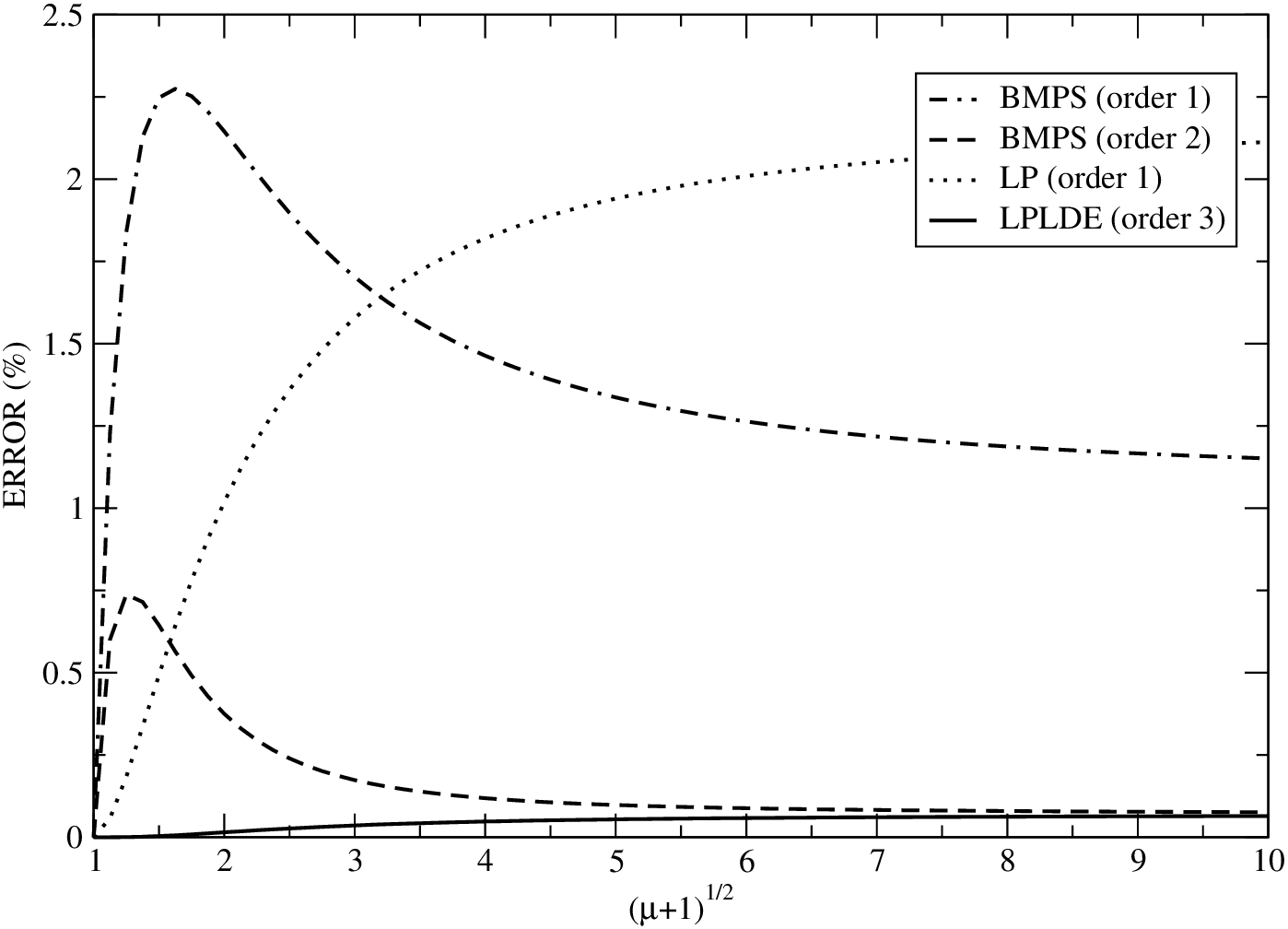}
\caption{Error corresponding to the different approaches for the case studied in Fig.~3}
\end{figure}

\end{document}